\documentclass[journal=jacsat,manuscript=article]{achemso}

\usepackage[version=3]{mhchem} 

\author{Daniel Schmidt}
\affiliation{TXproducts UG (haftungsbeschr{\"a}nkt), Luruper Hauptstr. 1, 22547 Hamburg, Germany}
\author{David von Stetten}
\affiliation{European Molecular Biology Laboratory (EMBL), Notkestrasse 85, 22607 Hamburg, Germany}
\author{Michael Agthe}
\affiliation{European Molecular Biology Laboratory (EMBL), Notkestrasse 85, 22607 Hamburg, Germany}
\author{Daniel Hensel}
\affiliation{Leibniz-Institut f{\"u}r Kristallz{\"u}chtung, Max-Born-Str. 2, 12489 Berlin, Germany}
\author{Arwen R. Pearson}
\affiliation{University of Hamburg, Luruper Chaussee 149, 22761 Hamburg, Germany}
\author{Goddfrey S. Beddard}
\affiliation{School of Chemistry, University of Edinburgh, David Brewster Road, EH9 3FJ, UK}
\author{Briony A. Yorke}
\affiliation{School of Chemistry, University of Leeds, Leeds, UK }
\author{Friedjof Tellkamp}
\affiliation{Max Planck Institute for the Structure and Dynamics of Matter, Luruper Chaussee 149, 22761 Hamburg, Germany}
\author{Peter Gaal}
\affiliation{Leibniz-Institut f{\"u}r Kristallz{\"u}chtung, Max-Born-Str. 2, 12489 Berlin, Germany}
\alsoaffiliation[Company]
{TXproducts UG (haftungsbeschr{\"a}nkt), Luruper Hauptstr. 1, 22547 Hamburg, Germany}
\email{peter.gaal@ikz-berlin.de}
\title{Demonstration of Advanced Timing Schemes in Time-Resolved X-ray Diffraction Measurements}


\begin{document}

\section{Abstract}
We present time-resolved X-ray diffraction measurements using advanced timing schemes that provide high temporal resolution while also maintaining a high flux in the X-ray probe beam. The method employs on-off patterned probe pulse sequences that are generated with the \textit{WaveGate} solid-state pulse picker. We demonstrate the feasibility of our method at two different beamlines on millisecond and microsecond timescales.

\section{Introduction}
\label{sec:intro}

Next-generation synchrotrons\cite{Shin2021a} present exciting opportunities for researchers across various fields, including biology, chemistry, life sciences, materials science, and engineering\cite{Chap2023a}. A key advantage of these new sources is their improved ability to observe samples \textit{in situ} within their natural environments and monitor their dynamics \textit{in operando}. The relevant timescales for detecting structural dynamics typically range from picoseconds to milliseconds. Different methods exist at synchrotrons that can resolve these transient effects, e.g. X-ray Photon Correlation Spectroscopy (XPCS). The highest temporal resolution can be reached with the pump-probe method where a short stimulus, such as a laser pulse, excites the sample, and a subsequent probe pulse detects the transient state at a specific pump-probe delay. Usually, each delay point is measured multiple times to assure reliability of the measured data. The temporal resolution of this scheme is generally limited by the duration of the probe pulse, typically around 100\,ps\ at a synchrotron source \cite{Gaal2012a, Enqu2010a}. At a synchrotron, using the X-rays from only a single electron bunch is usually the shortest possible probe, but in many cases does not deliver sufficient signal. The most common method to isolate single pulses from the emitted synchrotron pulse train is external gating of the X-ray detector\cite{Shay2017}. However, sample and measurement equipment are still exposed to the X-ray beam throughout the whole excitation cycle, which may cause saturation effects or even damage to the sample or to the measurement equipment \cite{klur2024a}. Thus, when using the pump-probe method, researchers face a trade-off between temporal resolution and the sensitivity of their experiments.

In this article, we discuss advanced timing schemes that overcome these limitations to some degree. Specifically, we present a method that employs patterned probe sequences derived from a Hadamard transform. The Hadamard transform is already applied in spectroscopy\cite{Yong2025a,kupc2003a} and microscopy\cite{Rous2012a, Couf1982a} to maintain spectral or spatial resolution while increasing a measurement's sensitivity\cite{Nels1970a,bedd2009a}. Researchers have already adapted the Hadamard method to the time domain successfully by imaging a rotating disc with an X-ray beam \cite{Cao2009a}, capturing protein structural changes by time-resolved X-ray diffraction \cite{York2014a} and pump-probe photothermal deflection \cite{bedd2016a}. These results demonstrate the potential to significantly improve time-resolution, however, Hadamard time-domain measurements are not commonly performed due to the lack of flexibility in patterning the time structure of the X-ray beam. We employ a solid-state pulse picker (\textit{WaveGate}, TXproducts) to overcome this constraint.

\section{Employing a Hadamard Transform for Time-Domain Measurements}
\label{sec:hada}

\begin{figure}[htbp]
\centering
	\fbox{\includegraphics[width = \textwidth]{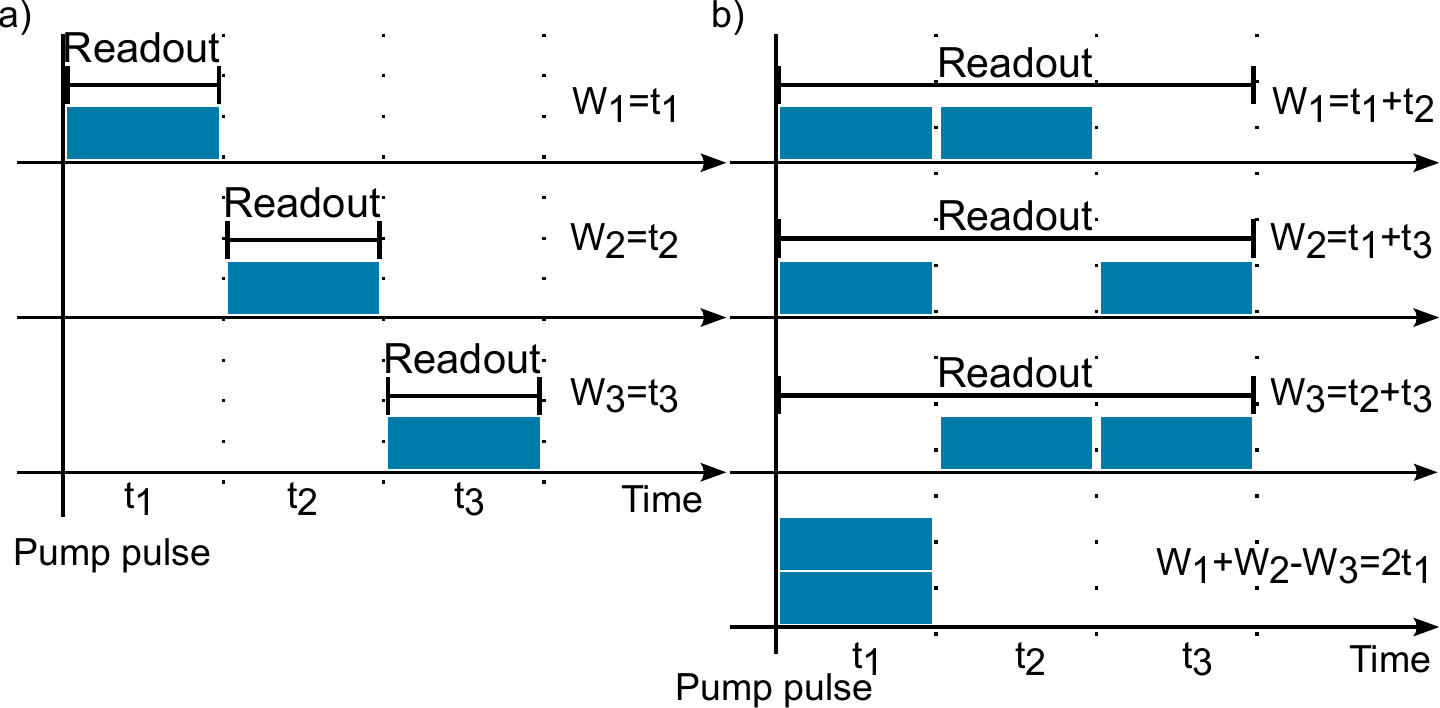}}
	\caption{\textbf{Time-resolved measurement schemes:} a) Standard pump-probe method: the delay axis is divided into time windows which are filled with probe pulses in consecutive measurements. Each measurement captures the transient state at one delay. b) Hadamard method: multiple windows on the delay axis are recorded with each probe sequence. The transient state at each individual delay time (only $t_1$ is shown) can be retrieved during post processing by combining all measurements.}
	\label{fig:GraphicalHadamard}
\end{figure}

In this section we will briefly describe the working principle of the Hadamard transform specifically in the context of time-domain measurements. We compare our method to pump-probe time-resolved measurements and show that it can yield the same temporal resolution even when patterned pulse sequences are used. Let $t$ denote the pump-probe delay axis. The pump-probe measurement divides this axis in $n$ discrete time windows called delays $t_{n}$. For simplicity, assume that all $t_{n}$ have the same duration $\delta t_{n}$ although this is not a necessary condition. In the pump-probe scheme, each time window is filled with a probe pulse and a measurement, \textit{e.g}., a detector image $W$ is recorded for each delay. The duration of the time window determines the temporal resolution. In a real experiment, the window could be filled either with only one or with several synchrotron pulses. We only require a unique pulse shape in all time windows. The full dataset contains $W_{n}$ measurements or detector images which represent the transient response $R(t_{n})=R_{n}$ of the sample. Figure~\ref{fig:GraphicalHadamard}~a) depicts the pump-probe scheme in a simple scenario where the delay axis $t$ is split into $n=3$ time windows. Each column in Figure~\ref{fig:GraphicalHadamard}~a) depicts a probe sequence, which in turn contains only one probe pulse. This means that only one time window is probed in each measurement. These probe sequences are represented by the following delay or \textbf{M}-matrix:
\begin{equation}
    \textbf{M} = \begin{bmatrix}
        1 & 0 & 0 \\
        0 & 1 & 0 \\
        0 & 0 & 1 \\
    \end{bmatrix}
\end{equation}

This measurement matrix (which is an identity matrix) represents the time-points in a traditional pump-probe experiment. Each detector image is obtained by multiplying a row of the \textbf{M}-matrix with the transient sample response $\textbf{R}$, and $\textbf{W}=\textbf{M}\times \textbf{R}$. The transient sample response is given by the inverse operation $\textbf{R}=\textbf{M}^{-1}\times \textbf{W}$. 

Consider now a different scheme for time-resolved measurements that employs probe sequences with more than one time window used. The scheme is sketched in Figure~\ref{fig:GraphicalHadamard}~b). Intuitively, it may seem that this measurement cannot yield the same temporal resolution as the previously described pump-probe method. This is however not the case.

To demonstrate this, we define a \textbf{S}-matrix which contains the probe sequences. Again, we choose a simple example with $n=3$ time windows. The \textbf{S}-matrix is derived from a Hadamard matrix\cite{York2014a, Sloa1982a, harw1979a} and has the form
\begin{equation}
    \textbf{S} = \begin{bmatrix}
        1 & 1 & 0 \\
        1 & 0 & 1 \\
        0 & 1 & 1 \\
    \end{bmatrix}
\end{equation}
The first measurement yields $\textbf{W}_{1}=\textbf{S}_{1}\times \textbf{R}=(1\cdot \textbf{R}_{1})+(1\cdot \textbf{R}_{2})+(0\cdot \textbf{R}_{3})$, \textit{i.e}., it is a summation over a total of two probe pulses in windows 1 and 2. Again, we want to obtain the sample response $R$, thus we perform the inverse operation $\textbf{R}=\textbf{S}^{-1}\times \textbf{W}$, where 
\begin{equation}
    \textbf{S}^{-1} = 0.5\cdot \begin{bmatrix}
        1 & 1 & -1 \\
        1 & -1 & 1 \\
        -1 & 1 & 1 \\
    \end{bmatrix}
\end{equation}
For the first time window of the transient response we find $\textbf{R}_{1}=\textbf{S}_{1}^{-1}\times \textbf{W}=0.5\cdot (\textbf{W}_{1} + \textbf{W}_{2} - \textbf{W}_{3})$. It is immediately evident that full measurement of all probe sequences is necessary to retrieve the transient state $\textbf{R}_{1}$. The same is true for any other transient state at any other delay. Figure~\ref{fig:GraphicalHadamard}~b) depicts graphically how the first delay point can be retrieved.

To conclude this section, we want to point out the following general remarks on time-domain Hadamard measurements: First, the Hadamard method can provide the same time resolution as the pump-probe method. Second, to retrieve the time-resolved response in a measurement one must first measure a full dataset of probe sequences and extract the dynamic response through the inverse operation $\textbf{R}=\textbf{S}^{-1}\times \textbf{W}$. 

The advantage of the Hadamard method is the increased sensitivity while maintaining the temporal resolution. Specifically, it allows the probing of fast processes with intense probe pulses that encompass multiple delay windows. A necessary requirement for using the Hadamard method is the ability to encode the complex sequences into the probe beam. We will present a solution to this challenge in the following section.

\section{Generating Hadamard Sequences in the Time Domain}
\label{sec:WaveGate}
Advanced timing schemes like the one discussed in the previous section are not commonly used in X-ray science. The reason is the lack of methods to easily modulate the time structure of synchrotron beams to deliver the complex, irregular Hadamard time sequences. We have recently presented a full characterization of a programmable solid state pulse picker that can achieve exactly this \cite{Schm2024a}. The device is called \textit{WaveGate}. Here, we will only briefly introduce its working principle and list the main performance parameters.

\begin{figure}[htbp]
	\centering
	\fbox{\includegraphics[width = 0.5\textwidth]{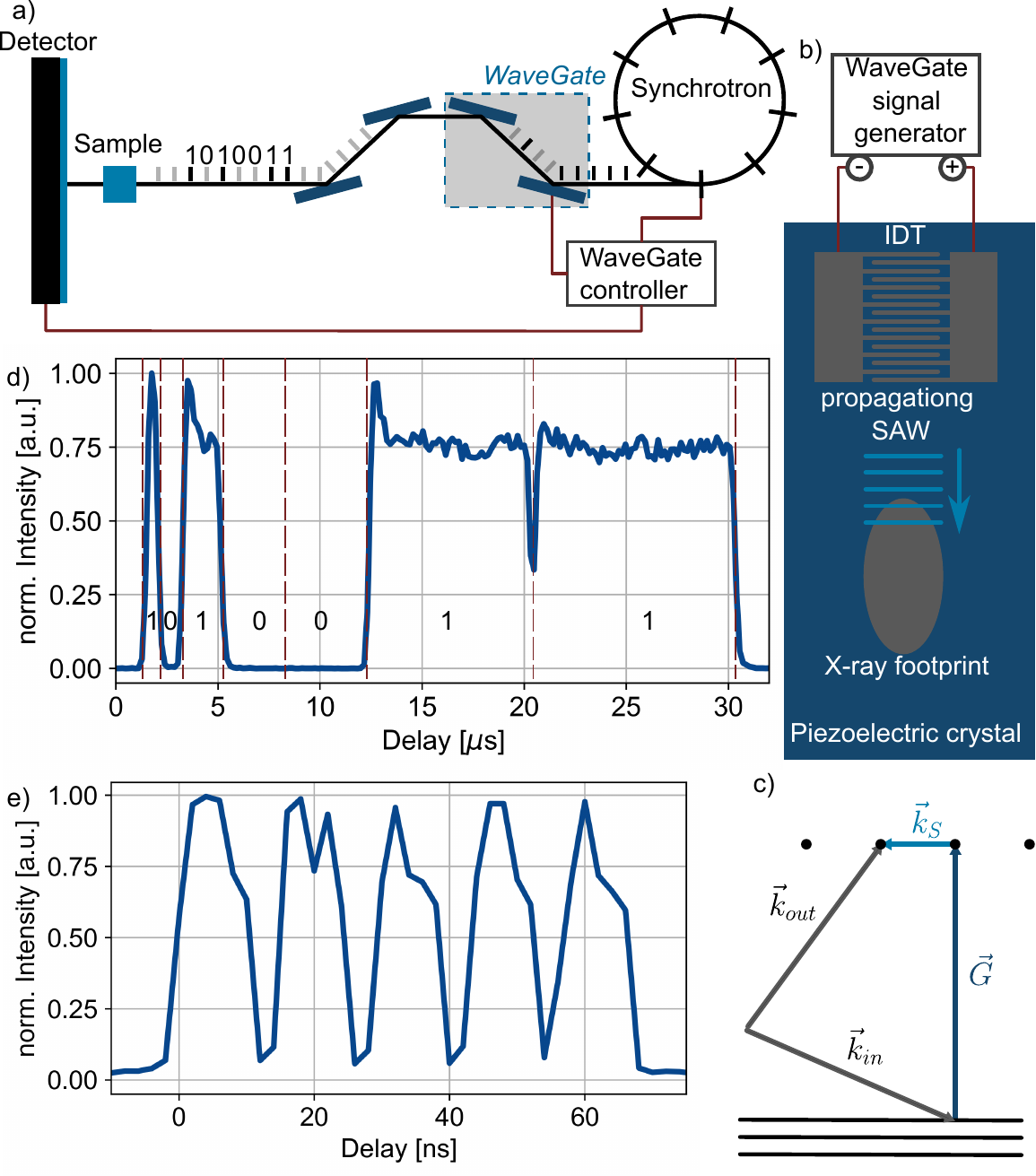}}
	\caption{\textbf{Experimental technique:} a) General beamline setup for synchrotron-based experiments using the \textit{WaveGate} pulse picker. X-ray pulses are emitted from a synchrotron storage ring and impinge on the \textit{WaveGate} setup\cite{Schm2024a}. The subsequent two crystals compensate for the change in beam height introduced by the \textit{WaveGate} and were used in the experiment discussed in Section~\ref{sec:LinearMotion} but not in the one in Section~\ref{sec:fastexp}. b) Top view on the active crystal of the \textit{WaveGate} pulse picker. The interdigital transducer (IDT) is exposed to a RF signal which is converted into a surface acoustic wave (SAW) by the piezoelectric substrate. The SAW modulates the diffraction efficiency of a structural Bragg reflex, thus enabling the transmission or suppression of the incident X-ray pulses. c) Laue condition of the active crystal. The device is discussed in detail elsewhere\cite{Schm2024a}. d) Exemplary probe pulse sequence with 7 time windows of variable duration: 700~ns, 1000~ns, 2000~ns, 3000~ns, 4000~ns, 8000~ns, 10000~ns. The individual windows are separated by a red dashed line, and the binary value of each matrix element is indicated. e) Isolation of single bunches from the 5x9 bunch pattern at CHESS by scanning with a 10~ns \textit{WaveGate} window.}
	\label{fig:Setup}
\end{figure}

The $\textit{WaveGate}$ consists of two crystalline piezoelectric substrates that are mounted in a double monochromator configuration, as shown in Figure~\ref{fig:Setup}~a). The first, so-called active crystal is depicted in Figure~\ref{fig:Setup}~b). An interdigitated transducer (IDT) connected to a signal generator is the active component. The time-dependent electric field which is applied to the crystal via the IDT is converted to a time-dependent strain field. The shape and the characteristic dynamics of the strain field are determined by the geometry of the IDT and by the external control signal from the signal generator. Converting  electronic signals into sound waves in piezoelectric crystals is an important application of such materials, e.g., in mobile communications\cite{Izha2025a,Yang2023a} or in sensors\cite{Zhang2023a,Four2024a}. 

The WaveGate adapts this technology to realize novel functionalities in X-ray photonics by using the generated strain fields for manipulation of the diffraction efficiency of structural Bragg reflexes. In short, the strain field propagates as a bulk acoustic wave (BAW)\cite{Izha2025a} below the IDT and as a surface acoustic wave (SAW)\cite{Ji2023a} in lateral direction along the surface of the crystal. The WaveGate can use both configurations. The wavevector associated to the periodic strain is $\vec{k_s}$. In combination with the structural Bragg reflex of the piezoelectric substrate with a reciprocal lattice vector $\vec{G}$ the X-ray beam diffracts into artificial satellites. The corresponding Laue condition $\vec{k_{in}}-\vec{k_{out}}=\vec{k_{s}}+\vec{G}$ is depicted in Figure~\ref{fig:Setup}~c) where $\vec{k_{in}}$ and $\vec{k_{out}}$ denote the wave vector of the incident and diffracted X-ray beam, respectively. 

A detailed explanation of the functionality and specification of the \textit{WaveGate} are presented elsewhere\cite{Schm2024a}. To manipulate a train of X-ray pulses into Hadamard sequences, the WaveGate must be synchronized to the bunch marker signal of the synchrotron storage ring. The timing jitter of the relevant signals is typically 100\,ps\cite{Hoso2001a,Evai2016a}, i.e., much smaller than the fastest switching time of the WaveGate. We have performed an extensive characterization of the WaveGate device\cite{Schm2024a}: The peak diffraction efficiency is 30\% of the incident beam. The on-off switching time can be as short as 2\,ns and the suppression of photons in the off-state lies between 10$^{-3}$ and 10$^{-4}$, depending on the configuration of the WaveGate. Admittedly, there is only few data on the long-term stability of the device. We have performed overnight loops of rocking curve scans of Bragg reflexes which show no variation in the position or the intensity of the measured reflex. In our experience, the device needs quick realignment every 1-2 days. However, it is unclear if drifts in the WaveGate setup or changes in the pointing of the incident synchrotron beam make the realignment necessary. So far, we have not observed a degradation of the active crystals in the X-ray beam. 

We demonstrate the capabilities of the WaveGate by isolating single bunches from the 9x5 bunch pattern at the CHESS 7B2 beamline as shown in Figure~\ref{fig:Setup}~e). We used an Eiger2 detector with a 100~ns gate to isolate a whole bunch train with 5 pulses separated by 14~ns. To isolate a single pulse we scanned a 10~ns temporal window spanned by the WaveGate with respect to the bunch marker.

An exemplary Hadamard sequence generated by the \textit{WaveGate} with $n=7$ time windows is plotted in Figure~\ref{fig:Setup}~c). Note that, in this example, the duration $\delta t_{n}$ of the time windows $n$ is variable, \textit{i.e}., the duration of $t_{0} = 0$ is $\delta t_{0}$=700\,ns and the duration $\delta t_{6}$ is 10\,$\mu$s. Each time window is separated from its adjacent window by vertical dashed lines. The full Hadamard sequence reads 1010011. To modulate the time-dependent transmission of the X-ray beam through the \textit{WaveGate}, we trigger the \textit{WaveGate} signal generator with the same sequence and modulate each element $n$ with its appropriate duration $\delta t_{n}$ . The active trigger generates an output at the signal generator which translates into diffraction satellites as previously described. We provide TANGO\cite{goet2020a} and EPICS\cite{dale2020a} servers to incorporate the necessary hardware into BLISS\cite{delg2020a} or BlueSky\cite{alla2019a} or similar beamline control environments.

Figure~\ref{fig:Setup}~a) shows a schematic of the \textit{WaveGate} installed in a beamline. The experiments presented in the next section were performed at beamlines P14-EH2 (T-REXX)\cite{IUCr2019a} and P23 at PETRA III (DESY, Hamburg). We used a bunchmarker signal to synchronize the \textit{WaveGate} to the bunches in the storage ring. However, for time windows much larger than the bunch separation, the synchronization can be omitted. In this case, the \textit{WaveGate} delivers a trigger signal to synchronize the rest of the experiment to the Hadamard time sequence.

\section{Monitoring Mechanical Motion with Hadamard Time Domain Measurements}
In the following we demonstrate the ability to resolve transient dynamics using the advanced timing scheme discussed in the previous sections. We performed two measurements to exemplify the method: first, we monitor the back-and-forth motion along the X-ray beam axis of a translation stage using Debye-Scherrer diffraction from a powder sample and second, we monitor the micromechanical oscillation of a 500\,nm thin SiN membrane after absorption of a laser pulse by Laue diffraction.  

In both cases we employ a $7\times 7$ \textbf{S}-matrix that is displayed in Equation~\ref{equ:hada7}.

\begin{equation}
\label{equ:hada7}
    \textbf{S}_7=\begin{bmatrix}
    1 & 1 & 1 & 0 & 1 & 0 & 0 \\
    1 & 1 & 0 & 1 & 0 & 0 & 1 \\
    1 & 0 & 1 & 0 & 0 & 1 & 1 \\
    0 & 1 & 0 & 0 & 1 & 1 & 1 \\
    1 & 0 & 0 & 1 & 1 & 1 & 0 \\
    0 & 0 & 1 & 1 & 1 & 0 & 1 \\
    0 & 1 & 1 & 1 & 0 & 1 & 0
    \end{bmatrix}
\end{equation} 
 For completeness, the inverse matrix \textbf{S}$^{-1}$ is generated by Equation~\ref{equ:invHada7}\cite{bedd2009a}.
\begin{equation}
\label{equ:invHada7}
\textbf{S}^{-1} = 2\left(2\textbf{S}^{T}-\textbf{J}\right)/\left(n+1\right),
\end{equation} 
where \textbf{J} is an all-ones matrix and the superscript $T$ denotes the transpose.

\subsection{Back-and-forth motion of a linear translation stage along the X-ray beam axis monitored by Debye-Scherrer diffraction}
\label{sec:LinearMotion}

\begin{figure}[htbp]
	\centering
	\fbox{\includegraphics[width = 0.5\textwidth]{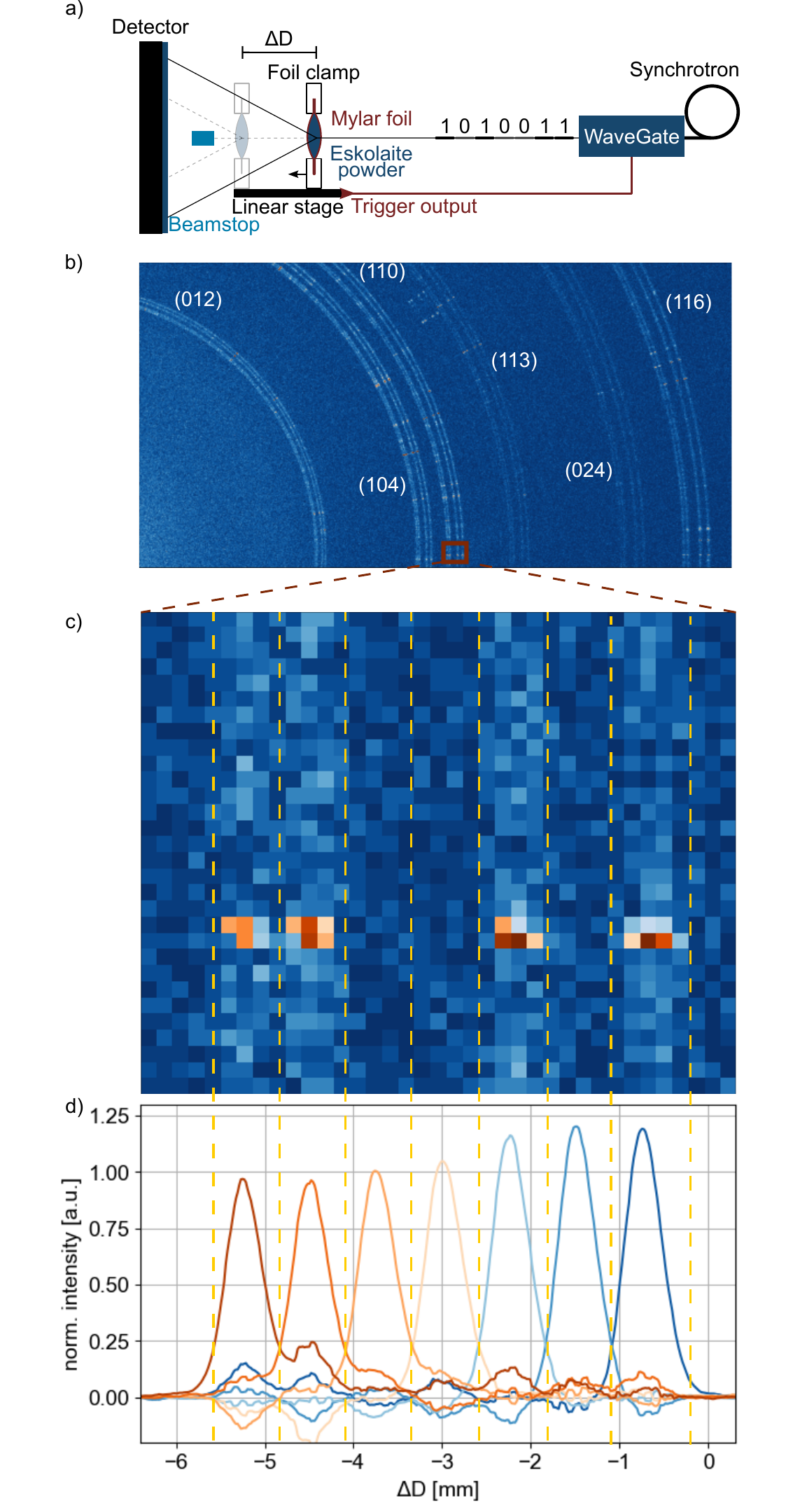}}
	\caption{\textbf{First Advanced Timing Measurement:} a) Experimental Setup: an Eskolaite powder sample is attached to a linear stage that performs a periodic back-and-forth motion along the X-ray beam axis. Incident X-rays diffract from the sample in Debye-Scherrer rings with a radius depending on the transient position of the translation stage as the sample to detector distance was varied. The sample was probed with a probe sequence generated by the \textit{WaveGate} pulse picker. b) Sum of 100 detector images depicting multiple Debye-Scherrer diffraction rings from the sample. The rings display a motion blur with a modulated intensity. c) Detailed view of the (110) Debye-Scherrer ring. The intensity modulation displays the inverted (from right to left) probe sequence 1010011. d) Transient stage position, obtained after azimuthal integration and the inverse transform with the full dataset.}
	\label{fig:Powder}
\end{figure}

This first demonstration measurement was performed at the T-REXX endstation on beamline P14 at the PETRA III storage ring and the experimental setup is depicted in Figure~\ref{fig:Powder}~a). The general beamline layout with the installed \textit{WaveGate} pulse picker was already discussed in the previous section~\ref{sec:WaveGate}. The X-ray beam with the modulated pulse sequences impinges a $\text{Cr}_2\text{O}_3$ Eskolaite powder sandwiched between two Mylar foils. The sample was mounted on a linear translation stage (SmarAct) that performed a repetitive back-and-forth motion along the X-ray beam axis $\Delta$D. The incident beam had a photon energy of 12.7\,keV. It diffracted into several Debye-Scherrer rings that were captured by an area detector (Eiger 4M, Dectris). Figure~\ref{fig:Powder}~b) depicts a section of a sum of 100 detector images. The area around the (110) ring marked by the red square is magnified in Figure~\ref{fig:Powder}~c). 

The synchronization and timing scheme that was implemented for this measurement was as follows. The $n=7$ time windows had an equal duration of 100\,ms. For this demonstration we choose a duty cycle of 50\% to be able to distinguish consecutive filled time windows visibly on the detector [c.f. Figure~\ref{fig:Powder}~b) and c)]. Hence, a logical 1 is represented by a 50\,ms X-ray pulse followed by a 50\,ms dark period. A logical 0 is represented by a 100\,ms dark period. The translation stage provided a trigger signal that was used to synchronize the probe pulse sequence generated by the \textit{WaveGate}. After the 7-window sequence completed, the \textit{WaveGate} blocked the incident X-ray beam from the storage ring. Images were synchronised to always start at the same stage position, with an overall exposure time of 1000\,ms.

Figure~\ref{fig:Powder}~b) displays a blurring\cite{Hark2024a} of the Debye-Scherrer rings that stems from the mechanical motion of the Eskolaite powder sample mounted on the translation stage. In the magnified area displayed in Figure~\ref{fig:Powder}~c) we recognize the temporal probe sequence 1010011 (right to left). This sequence corresponds to the third column of the \textbf{S}-matrix displayed in Equation~\ref{equ:hada7} and the image represents measurement W$_{3}$ according to the formalism laid out in Section~\ref{sec:hada}. With the full set of images W$_{0}$ to W$_{6}$ we perform the inverse transformation $R=S^{-1}\times W$ to find the position of the powder ring on the detector and thereby the transient position of the translation stage in each time window $t_{n}$. The result is depicted in Figure~\ref{fig:Powder}~d), where early delay windows are coded in blue and later delays are coded in red. 

By employing the inverse transformation we can determine the individual transient positions of the stage even though the measured detector images are blurred due to the moving sample. The measured intensity at the different positions is reduced at later delays and we observe a smearing of the position in the corresponding traces. We suspect that the speed of the translation stage varied due to friction, which translates into slight position errors in the inverse transformation. This is a good example to point out the challenges and intricacies of this method. Because the probe pulses are distributed across many time windows, it is impossible to repeat a measurement of a single delay. Any errors spread to all delay times and if in doubt, the full series must be repeated.

\subsection{Micromechanical motion of a 500\,nm thin membrane monitored by Laue diffraction}
\label{sec:fastexp}

\begin{figure}[htbp]
	\centering
	\fbox{\includegraphics[width = \textwidth]{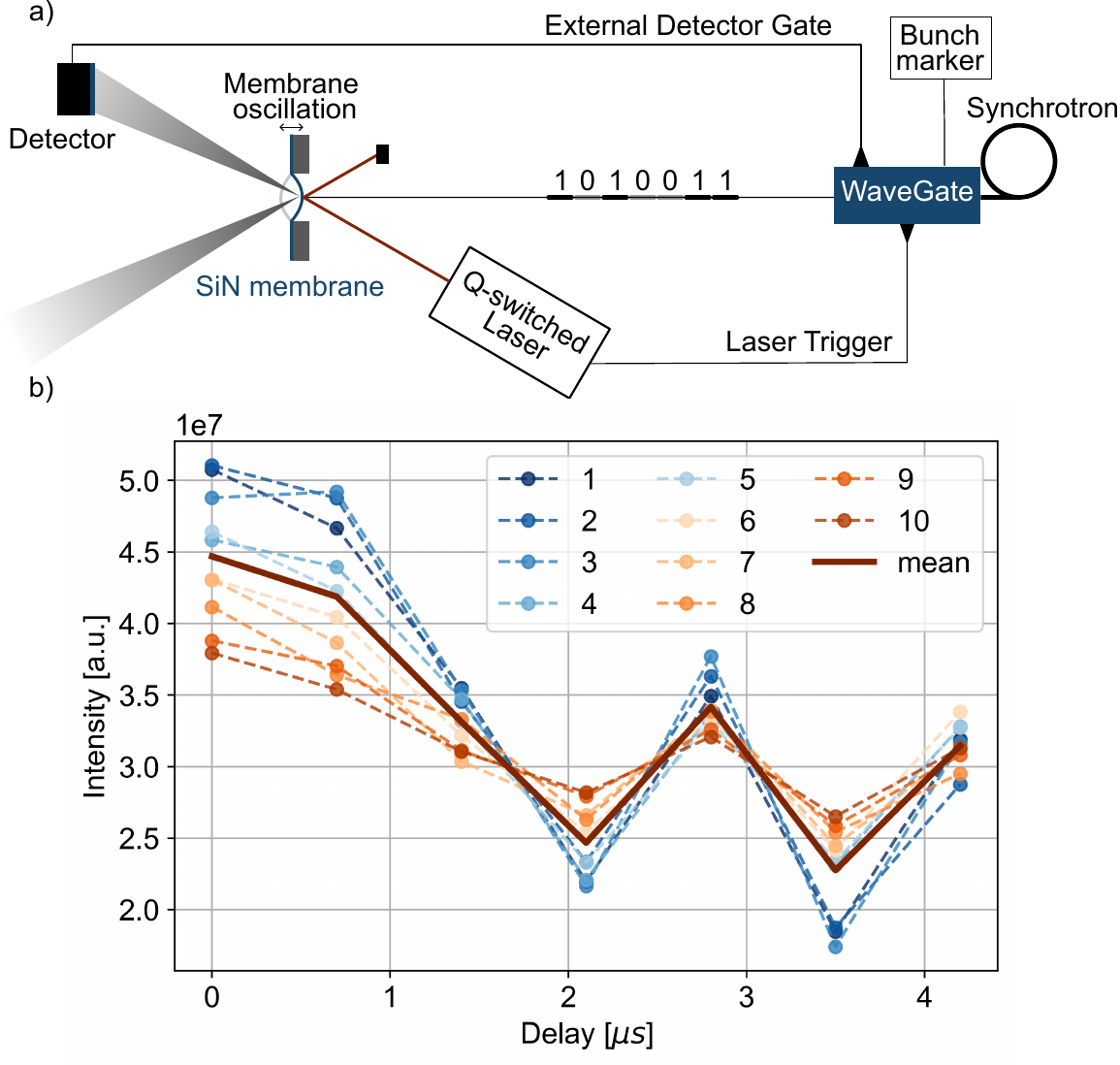}}
	\caption{\textbf{Second Advanced Timing Measurement:} a) Experimental setup: the \textit{WaveGate} pulse picker was synchronized to the bunchmarker signal from the storage ring and generated trigger signals for the excitation laser and area detector. The sample (SiN membrane with 40\,nm Pt coating) was excited by 7\,ns laser pulses with a wavelength of 1064\,nm. The subsequent dynamics were probed by alternating sequences of probe pulses generated by the \textit{WaveGate} pulse picker. Incident X-rays were diffracted in Debye-Scherrer rings of which only a small fraction was captured by the area detector. b) Transient sample dynamics obtained from the inverse transformation using all recorded probe sequences. Each sequence was recorded ten times (bullets, dashed) to demonstrate the reliability of the measurement. The average response is depicted in the red solid line. Laser-induced degradation of the sample leads to a decay of the signal as the measurement progresses.}
	\label{fig:Membrane}
\end{figure}

We now discuss a second, more realistic demonstration measurement, where the transient response of the sample was not known in advance. The sample was a 500\,nm thin SiN membrane which was coated with an amorphous 40\,nm thin Pt film by magnetron sputtering. The experimental setup was very similar to the scheme depicted in Figure~\ref{fig:Setup}~a). We performed this experiment at beamline P23 at the PETRA III storage ring and a schematic is given in Figure~\ref{fig:Membrane}~a). The sample was mounted in the direct beam diffracted from the second \textit{WaveGate} crystal. The X-ray beam was moderately focused to a spot size of 30\,$\mu$m $\times$ 160\,$\mu$m by a set of Be compound refracting lenses located 30 m upstream from the sample position. We captured the (100) diffraction ring in Laue geometry with an area detector (Lambda 750k, X-Spectrum). The distance between sample and detector was 1\,m and the diffracted beam from the sample covered the entire detector area of 28 $\times$ 85\,mm$^{2}$. With the available detector area we captured only a small fraction of the total radiation diffracted from the sample.

To elicit a transient response, we exposed the sample to laser pulses with a duration of 7\,ns, a wavelength of 1064\,nm and a maximum pulse energy of 1\,mJ. The laser footprint at the sample position was almost circular, with a radius of 500\,$\mu$m. The sample was excited at a repetition rate of 1\,kHz with an incident fluence of 40.7\,mJ/cm$^{2}$.

The storage ring operated in 40 bunch mode, \textit{i.e.}, subsequent X-ray pulses were separated by a gap of 192\,ns. By reducing the repetition rate of the experiment to the 1\,kHz repetition frequency of the laser we reduced the X-ray flux by a factor of 5200\cite{Gaal2023}. Again, we used the 7$\times$7 \textbf{S}-matrix given in Equation~\ref{equ:hada7}, this time with a duration $\delta$t of each time window of 700\,ns. The \textit{WaveGate} was synchronized to the bunchmarker signal. A trigger signal for the laser and an external gating signal for the area detector were generated by the $\textit{WaveGate}$ electronic hardware. The external gating was active for the total duration of the probe pulse sequence. We tuned the delay axis such that the first delay point t$_{0}$ coincided with the arrival of the laser pump pulse at the sample. Therefore, no negative delay times were measured. For every probe pulse sequence we captured ten redundant images. Thus, for each of the 7 Hadamard sequences we performed a total of 10 delay scans over a delay range of 4.9\,$\mu$s. 

The transient sample response for the ten independent measurements was obtained through the inverse transform $\textbf{R}=\textbf{S}^{-1}\times \textbf{W}$. Results are depicted in Figure~\ref{fig:Membrane}~b) in the colored dashed lines. The dark red solid line shows the average of all measurements. We observe a continuous decay from the first to the last measurement, which indicates a laser-induced degradation of the sample. Otherwise, all ten measurements show a qualitatively identical temporal response. At this point we refrain from speculating about the underlying physical mechanism responsible for the observed dynamics. The only purpose of this measurement is to demonstrate the reliability of the advanced timing scheme. 

\section{Use cases for the Hadamard Timing Scheme}

\begin{figure}[htbp]
	\centering
	\fbox{\includegraphics[width = 0.8\textwidth]{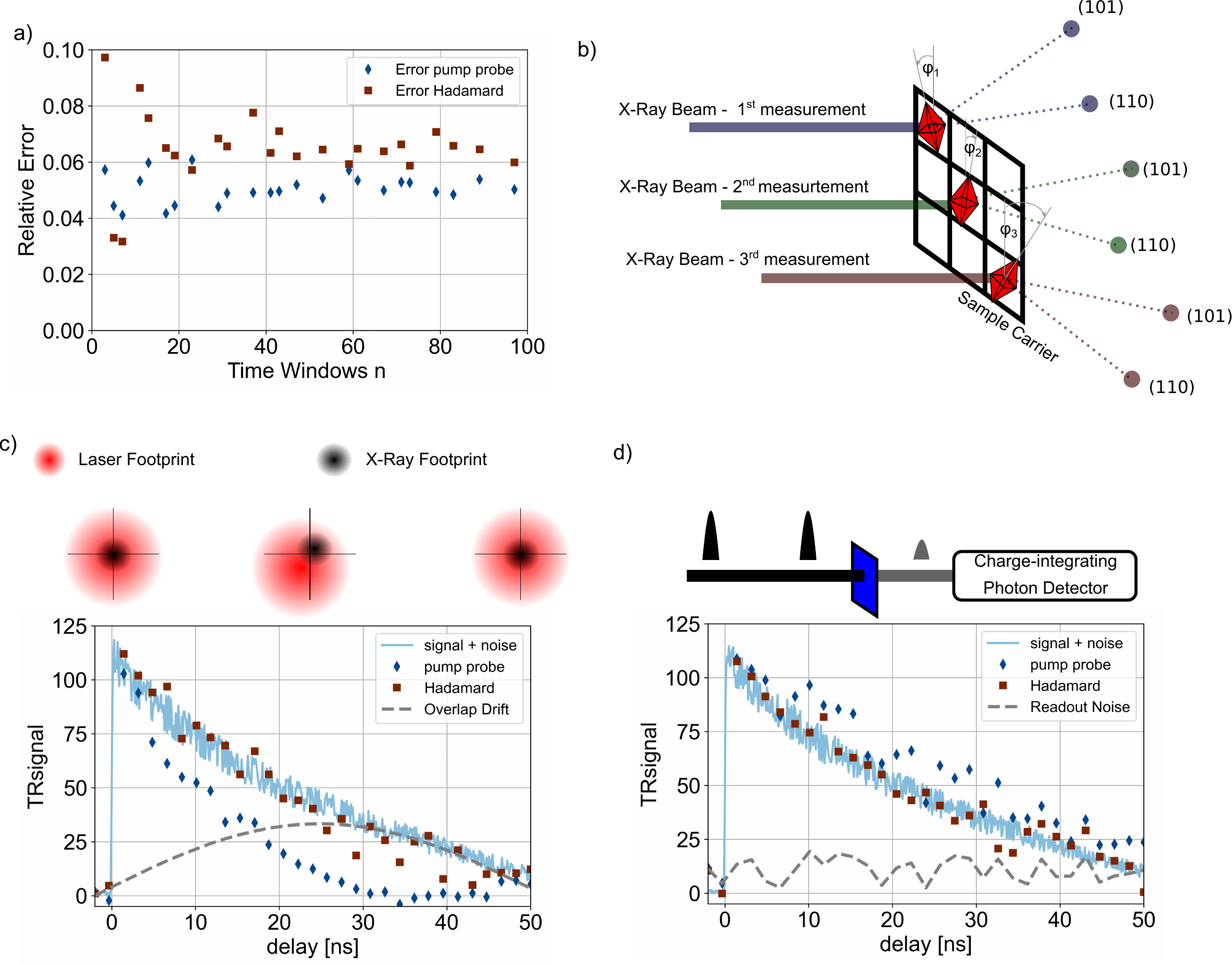}}
	\caption{\textbf{Application Examples for the Hadamard Timing Scheme:} a) Comparison of the signal-to-noise ratio of pump-probe (blue diamonds) and Hadamard (red squares) time-resolved measurements. The pump-probe method proves slightly superior to Hadamard timing. b) Serial protein crystallography measurement: a sample carrier contains multiple crystals with random orientation which are sequentially put into the X-ray beam. The measurement cannot be performed in a direct pump-probe scheme due to the insufficient photon count for orientation indexing. The Hadamard method can mitigate these limitations (see main text for details). c) and d) influence of experimental instabilities, e.g. thermal drift of the spatial overlap between pump and probe beam (c) or additional readout noise from the charge-integrating detector (d). The Hadamard measurement (red squares) reproduces the original dynamics while the pump-probe measurement (blue diamonds) produces false results.}
	\label{fig:ErrorComp}
\end{figure}

In this section, we explore the application of the Hadamard timing scheme in time-resolved X-ray experiments. We begin by estimating the signal-to-noise ratio (SNR) in an idealized scenario, where a Poisson-distributed stream of X-ray photons is detected by an ideal detector. For a given signal intensity $I_0$, the expected measured pump-probe signal level is $I_0\pm\sqrt{I_0}$,  where the latter term represents the standard deviation due to photon counting statistics. 

Quantifying the measurement error in a Hadamard-based acquisition is less straightforward. The total intensity for a single Hadamard illumination is given by $I_H = I_0\times(n/2 + 0.5)$, with an associated uncertainty of $\sqrt{I_H}$. This measurement is repeated $n$ times, and the resulting errors are distributed across $n$ time windows via the inverse Hadamard transform.

To compare the SNR of both measurement schemes we simulated an idealized sampling experiment. The input signal was set to a constant amplitude $I=100$. The relative measurement error in each time window was then computed either directly (pump-probe) or after applying the inverse Hadamard transform. Simulations were performed for all valid Hadamard sequences with window sizes $n<100$, i.e., all prime numbers between $3$ and $97$. The average measurement error of each simulation is shown in  Figure~\ref{fig:ErrorComp}~a). The pump-probe measurement (blue diamonds) yields a consistent relative error of $\approx5\%$, independent of the number of time windows. In contrast, the Hadamard method (red squares) exhibits higher error for small $n$, but converges to a similar error level as the pump-probe method for $n>20$. Under ideal conditions, the Hadamard scheme is therefore slightly inferior in terms of SNR.

We next compare both timing schemes under non-ideal conditions, including low photon flux, experimental drift, and read-out noise in charge-integrating X-ray detectors. The first case study involves time-resolved serial protein crystallography (Figure~\ref{fig:ErrorComp}~b), as detailed in Ref.~\cite{York2014a}. Due to the limited lifetime of protein crystals in the X-ray beam, data must be collected from multiple crystals, each randomly oriented and sequentially introduced into the beam. The pump-probe method suffers from two major limitations at high temporal resolution: (1) Time resolution is achieved via external gating of the detector, meaning that most of the X-ray exposure does not contribute to usable data, thereby reducing the effective sample lifetime; and (2) insufficient photon counts for reliable orientation determination and indexing, preventing structural reconstruction. This limitation is mitigated in the Hadamard scheme, which maintains effective sample lifetime and enables indexing due to higher photon counts per recorded image.

The second case considers a laser-pump - X-ray-probe experiment affected by alignment drift of optical components, such as those caused by thermal fluctuations. We assume that the drift occurs on a timescale slower than the integration time of a single delay point. In the pump-probe method, such drifts alter the excitation conditions within the probe volume, leading to distorted dynamics. This scenario is simulated in Figure~\ref{fig:ErrorComp}~c). The original signal (light blue line) is distorted in the pump-probe measurement (blue diamonds) due to spatial drift (gray dashed line), potentially resulting in an overestimation of the system's equilibration time. In contrast, the Hadamard method is robust against such drift and accurately reproduces the original signal.

A similar scenario is presented in Figure~\ref{fig:ErrorComp}~d), where the probe measurement is degraded by read-out noise from a charge-integrating X-ray detector~\cite{Deng2023a,Beck2016}. This noise manifests as both an apparent delay in equilibration and additional random fluctuations. Again, the Hadamard method demonstrates superior fidelity in recovering the true signal dynamics and may, therefore, be suitable for application to a wide-range of pump-probe experiments where it is possible to modulate the probe according to sequences defined by the \textbf{S}-matrix. \textit{WaveGate} modulation is particularly suitable for serial synchrotron X-ray crystallography \cite{caramello2024femtoseconds, lewis2024small} or X-ray absorption spectroscopy \cite{chergui2017photoinduced} experiments investigating irreversible and dose sensitive processes in chemistry and biology, where averaging the signal from repeated measurement of the same sample volume is not possible.

In summary, our simulations show that the Hadamard timing scheme can outperform the pump-probe method in terms of accuracy and sensitivity under realistic experimental conditions. Its advantages become particularly evident in the presence of experimental instabilities such as thermal drift, electronic or mechanical noise, or nonlinearities in beamline components.  Conversely, the pump-probe method remains optimal when photon shot noise is the dominant source of uncertainty.

\section{Summary and Conclusion}
\label{sec:conclusion}
In conclusion we have demonstrated the implementation of advanced timing schemes for synchrotron-based time-resolved measurements by encoding of the incident X-ray beam. The new method employs probe sequences of probe pulses which are distributed along a delay axis according to a predefined rule derived from a Hadamard transform. By performing the inverse transform on the recorded data the transient response of a sample at each delay can be obtained. The main advantage of this timing scheme is its increased sensitivity and the increased photon flux during the measurement.

A prerequisite to employ the new timing scheme is the ability to generate different probe sequences with high accuracy and variability. Our solution to this challenge is the \textit{WaveGate} pulse picker. It is a versatile device that integrates easily in different beamline environments and allows the generation of probe patterns at the beamline control interface. 

We demonstrated the feasibility of our method with two exemplary measurements that were performed at different beamlines at the PETRA III storage ring at DESY, Hamburg.

\section{Acknowledgment} 
The T-REXX endstation and the development of the \textit{WaveGate} pulse picker for encoding Hadamard sequences are supported by the Bundesministerium für Bildung und Forschung (Verbundforschungsprojekte 05K16GU1, 05K19GU1 \& 05K22GU6). 

Beamtime at T-REXX was provided by the T-REXX BAG at EMBL (MX862). We acknowledge DESY (Hamburg, Germany), a member of the Helmholtz Association HGF, and EMBL for the provision of experimental facilities. Parts of this research were carried out at PETRA III and Cornell High Energy Synchrotron Source. We would like to thank Dr. Dmitry Novikov and Dr. Azat Khadiev for assistance in using Beamline P23 and Dr. Donald Walko and Dr. Stephen Paul Meisburger for assistance in using Beamline ID7b2. Beamtime was allocated for proposal I-20220569 and 4071-A, respectively.

\section{Disclosures} D.S. and P.G. are affiliated with TXproducts, i.e., the manufacturer of the WaveGate pulse picker. The authors declare no conflicts of interest.

\section{Data availability} Data underlying the results presented in this paper are not publicly available at this time but may be obtained from the authors upon request.

\providecommand{\latin}[1]{#1}
\makeatletter
\providecommand{\doi}
{\begingroup\let\do\@makeother\dospecials
	\catcode`\{=1 \catcode`\}=2 \doi@aux}
\providecommand{\doi@aux}[1]{\endgroup\texttt{#1}}
\makeatother
\providecommand*\mcitethebibliography{\thebibliography}
\csname @ifundefined\endcsname{endmcitethebibliography}
{\let\endmcitethebibliography\endthebibliography}{}


\end{document}